\documentclass{ws-p9-75x6-50}
\newcommand{\BEQ}{\begin{equation}}
\newcommand{\EEQ}{\end{equation}}
\newcommand{\BEA}{\begin{eqnarray}}
\newcommand{\EEA}{\end{eqnarray}}
\newcommand{\nn}{\nonumber}
\renewcommand{\d}{{\rm d}}

\newcommand{\ub}{\nu}

\begin{document} 
\title{
kinetic equation model for self-gravitating systems}
\author{A.E. Allahverdyan$^{1,2)}$ and Th.M. Nieuwenhuizen$^{3)}$}
\address{$^{1)}$ CEA/Saclay, Service de Physique Theorique, 
F-91191 Gif-sur-Yvette Cedex, France, \\
$^{2)}$Yerevan Physics Institute,
Alikhanian Brothers St. 2, Yerevan 375036, Armenia \\
$^{3)}$ Department of Physics and Astronomy,
University of Amsterdam, \\
Valckenierstraat 65, 1018 XE Amsterdam, The Netherlands }
\maketitle

\abstracts{
The kinetic theory of a self-gravitating system is considered in the
 Bhatnager-Gross-Krook approximation to  the kinetic equation. This
approach offers a unique and tractable setup for studying the central,
collision-dominated region of the system, as well as its almost
collisionless outer part.
Stationary non-equilibrium states of the system are considered and
several different self-similar solutions are identified.}


\section{ Introduction}
There is a long tradition of applications the kinetic theory to
the gas of self-gravitating particles \cite{chandra}-\cite{saslaw}.
The first regular approach of that kind was developed by Chandrasekhar 
\cite{chandra}, and appeared to be very useful in 
astrophysical applications. Indeed, the kinetic theory is
certainly applicable for globular star clusters,
which are rather well-isolated systems on the relevant
time-scales, and contain sufficiently many (typically of order
$N\sim 10^5$) stars \cite{shapiro}. 
Therefore, the kinetic approach, which typically
studies a closed, macroscopic system, has  a definite range of validity
here. It allows to identify correctly the collisional time-scale 
\cite{chandra}-\cite{saslaw}, and predict the moderate and late 
stages of the evolution for globular clusters \cite{MeylanHeggie}. 

The general method for constructing the collisional Fokker-Planck
equation for statistical systems with $1/r$ interaction was proposed
in \cite{judd} (see also the text-book descriptions in
\cite{chandra,landau,spitzer,saslaw,paddy}).  
This equation adopts a more general strategy for  
Boltzmann's kinetic equation to the considered long-range interacting
system. After suitably dealing with the peculiarities of the long-range
interaction, one ends up with the collisional Fokker-Planck equation.
The main difficulty with this equation is that it is technically rather
involved and resistant to reliable analytical treatments. 
Thus, one has to change
the level of description and move to hydrodynamical models of the
stellar dynamics (see e.g., \cite{spitzer,saslaw} and refs. therein). 
Instead of the one-particle distribution function, which is the basic
object of the kinetic theory, the hydrodynamical approach operates 
with a few coarse-grained hydrodynamic variables, but by its meaning 
it is applicable only to the dense central regions of the 
self-gravitating system, where collisions are certainly dominating. 

In this contribution we present another approach to the kinetic
theory of self-gravitating systems. This is an approach of kinetic
models, where instead of going immediately to the hydrodynamical level
of description, one is properly {\it modelling} the collisional part
of the kinetic equation, trying to preserve its main qualitative
characteristics, but to make the system analytically tractable. 
One of the most successful kinetic models was proposed by Bhatnager,
Gross and Krook (BGK) \cite{bgk,gross1,gross,grad}, 
and since that time appeared to be very
useful in the kinetic theory. The reason of its success is that the
model offers a minimal way to incorporate the necessary 
conservation laws (those of particle number, momentum and energy)
with a
simple relaxation mechanism. For a self-gravitating statistical 
system this approach provides natural methods to consider
simultaneously the collision-dominated (overdamped) regions of the
system, and its almost collisionless outer domains (underdamping). 

In the present contribution we will restrict ourselves to the
presentation of the BGK-method as applied to self-gravitating systems
(section II), and to straightforward analytical investigations which
will recover within a single setup several known and also new results 
for the steady regime of behavior (section III). More detailed 
investigation of the situation,  as well as the presentation of 
more elaborated BGK-type models, are planned for future.

\section{Bhatnager-Gross-Krook kinetic equation}
The statistical dynamics of $N$ point particles with unit mass interacting 
through Newton's inverse-square law, will be considered in 
the classical kinetic approach. This basically involves two important assumptions:
{\it (i)} The field acting on a given particle (test particle) can be represented as
a mean, self-consistent field plus a contribution from two-particle
collisions. {\it (ii)} As usual, a collision between a pair of
particles is taken to be independent of the others, and local 
(namely, it leads to sudden changes in the corresponding momenta, 
whereas the coordinates can be considered as fixed).

The mean-field assumption is reasonable for a system having 
long-range forces, since they suppress fluctuations.
On the other hand, the second assumption concerning the simple cumulative
effect of independent collisions was subjected to a certain criticism 
(see e.g., \cite{paddy}),
because the long-range unshielded gravitational coupling involve many
simultaneously interacting particles. Nevertheless, an impressive amount of
observational and numerical material has been obtained, which advocates the 
use of the present approach al least on certain time-scales 
\cite{spitzer,MeylanHeggie}.

According to our assumptions the contribution from the direct interaction
between the particles is added
on the right-hand side of the Liouville equation for the one-particle 
distribution function $f(\vec{r},\vec{v},t)$ as the corresponding 
collision integral ${\cal C}[f]$:
\begin{equation}
\label{1}
\frac{\d f}{\d t}\equiv
\frac{\partial f}{\partial t}+\vec{v}
\frac{\partial f}{\partial \vec{r} }-
\frac{\partial \phi}{\partial \vec{r} }\frac{\partial f}{\partial \vec{v} }
={\cal C}[f]
\end{equation}
The mass of each particle is taken be unity, $m=1$, 
and the potential $\phi (\vec{r})$ can be viewed as an 
external despite its self-consistent closure with the Poison equation:
\begin{equation}
\label{Poisson}
\Delta \phi = 4\pi G\int \d \vec{v}f(\vec{r},\vec{v},t)=4\pi G n(\vec{r})
\end{equation}
As usual in kinetic theory, the distribution function
$f(\vec{r},\vec{v},t)$ is normalized on the total number of particles,
namely
\BEA
\int \d \vec{v}\, f(\vec{r},\vec{v},t)=n(\vec{r})
\EEA
is the density of particles, and
\BEA
\int \d \vec{r}\, n(\vec{r})=N
\EEA
is the total number of particles.

Instead of deriving the collision integral of the kinetic 
equation from the underlying microscopic dynamics, 
Bhatnager, Gross and Krook proposed the
following model for the collisional integral \cite{bgk}:
\begin{equation}
\label{a36}
{\cal C}_{BGK}[f] =-
\nu (r) [ f (\vec{r},\vec{v},t) - f_0 (\vec{r},\vec{v},t) ],
\end{equation}
where $\nu (r)>0$, and $f_0$ is the best local Maxwellian
\begin{equation}
\label{a37}
f_0 (\vec{r},\vec{v},t)=\frac{n(\vec{r},t)}{(2\pi T(\vec{r},t))^{3/2}}
\exp \left (
-\frac{(\vec{v}-\vec{u}(\vec{r},t))^2}{2T(\vec{r},t)}
\right ), 
\end{equation}
where $n(\vec{r},t)$, $T(\vec{r},t)$, $\vec{u}(\vec{r},t)$ are defined by the
following conditions:
\begin{equation}
\label{a38}
n(\vec{r},t)=\int \d \vec{v}~f (\vec{r},\vec{v},t),
\end{equation}
\begin{equation}
\label{a39}
n(\vec{r},t)\vec{u}(\vec{r},t)=\int \d \vec{v}~\vec{v}~
f (\vec{r},\vec{v},t),
\end{equation}
\begin{equation}
\label{a40}
3n(\vec{r},t)T(\vec{r},t)=\int \d \vec{v}~(\vec{v}-\vec{u}(\vec{r},t))^2~
f (\vec{r},\vec{v},t).
\end{equation}
Due to this conditions the collision integral (\ref{a36}) conserves
probability, momentum
and energy, and in the spatially homogeneous case $T=T(\vec{r},t)$ it
enforces convergence of the distribution function $f
(\vec{r},\vec{v},t)$ to the corresponding Gibbs distribution.
One should notice that BGK collision integral (\ref{a36}) is still 
non-linear due to these self-consistency relations.
However, this non-linearity is based on the first three moments of $f$ only,
which means that BGK scheme will be friendly to moments equations and related
procedures. Another attractive property of BGK model is that though
being proposed {\it ad hoc}, it was later derived from the Boltzmann
equation in both rarefied (low-density), and dense limits 
\cite{gross}\cite{grad}. In particular, relations with more formal approximation
methods (e.g., those proposed by Chapman-Enskog or Grad) were
established \cite{gross1}\cite{grad}. Thus, there is a general 
expectation that this approximation will reasonably mimic the 
main properties of the original Boltzmann equation.

The general conditions of invariance and convergence do not specify $\nu (r)$,
which has a meaning of the inverse characteristic relaxation time. 
It should not depend on $\vec{v}$, since else the conservation laws 
cannot be satisfied. Obviously, the concrete choice of $\nu
(\vec{r},t)$ should be determined demanding that Eq.~(\ref{a36}) 
is as close as possible to the original Boltzmann equation. 
A standard assumption is that $\nu$ depends on the coordinate 
$\vec{r}$ and time $t$ through the $\vec{r}$-dependence of the 
first three moments, $\nu =\nu (n,\vec{u},T)$, and a general procedure 
was proposed to derive its concrete form \cite{bgk}. 
We will not go into details of that procedure, but we
will show that already rather simple qualitative considerations
determine its rough form, and then we will directly take for
it the expression for the inverse characteristic time, which was
obtained from the full Boltzmann equation in Refs.~\cite{chandra,spitzer}.
To determine $\nu $ one notices that in 
the rarefied case it is proportional to $n$, and then due to obvious 
dimensional reasons one has
\begin{equation}
\label{a41}
\nu (\vec{r},t)\propto \frac{n(\vec{r},t)}{T(\vec{r},t)^{3/2}}.
\end{equation}
In this context
one could wonder why only $T$ and not $\phi$ should enter $\nu$.
Since $\nu$ belongs to the collision integral, 
according to the common initial assumption the influence of the field is
not taken into account there. Indeed, in the full Boltzmann equation  
binary collisions are considered in such a way that only microscopic 
two-particle interaction enters there,
and not the mean-field. This is inevitable to find
tractable schemes, and means that 
collisions occur on such time and length-scales, where the effect of 
the mean-field is not relevant.

More precisely, we notice
the binary collision relaxation time obtained through the complete
Boltzmann equation \cite{chandra,spitzer,saslaw,paddy}
\BEQ 
t_r=\frac{0.065\langle v^2\rangle^{3/2}}{nmG^2\ln(0.4N)}
\EEQ
with the right hand side calculated at typical radius $R$, and where
we momentarily recovered the dimensional units.
Thus, we just take $1/\nu$ equal to the local relaxation time, so
\BEA 
&&\nu(r)=\frac{1}{t_r(r)}=\frac{nmG^2\ln(0.4N)}{0.065\langle
v^2\rangle^{3/2}}
\equiv \frac{15.4\,\ln(0.4N)\, G^2m^{5/2}n(r)}{T(r)^{3/2}}=
\gamma \frac{n(r)}{T(r)^{3/2}},\\
&&\gamma=15.4\,\ln(0.4N)G^2m^{5/2}
\EEA
is the inverse relaxation time at position $r$~\cite{spitzer}.
To estimate whether collisions are relevant compared with the motion in
the mean-field, one has to compare  $t_{rel}$ with the dynamical 
time-scale $t_d$. For a cluster with mass $M=Nm$ and typical radius
$R$ the dynamical time is 
\BEQ
\label{tdynlocal}
t_d=\frac{R}{\overline v}=
\frac{R^{3/2}}{\sqrt{GM}}
\EEQ
The situation $t_r(r) \ll t_d$ means that collisions dominate (overdamping),
and $t_r(r) \gg t_d$ indicates underdamping. This qualitative criterion will
be applied to determine the transition from the outer weakly-damped
part of the system (halo) to the inner strongly-damped part (core).

\subsection{H-theorem and convergence to equilibrium}
\label{H}
Let us now briefly show that if the conditions for equilibrium
are satisfied, then the BGK-collisional term (\ref{a36}) indeed 
induces convergence toward the Gibbs distribution. To this end we will 
consider the time-behavior of the Boltzmann-Gibbs entropy:
\BEQ
\label{BG}
S(t)=-\int \d \vec{r}~\d \vec{v}~f(\vec{r},\vec{v},t)\ln f(\vec{r},\vec{v},t), 
\EEQ
\BEQ
\label{BG1}
\dot{S}=\int \d \vec{r}~\d \vec{v}~\nu (r)[f-f_0]\ln f
+\int \d \vec{r}~\d \vec{v}~\left [
\vec{v}\frac{\partial f}{\partial \vec{r} }-
\frac{\partial \phi}{\partial \vec{r} }\frac{\partial f}{\partial \vec{v} }
\right ]\ln f
\EEQ
Using the fact that 
$\int \d \vec{v}~[f-f_0]\ln f_0 =0$, 
which is conservation of probability, momentum and energy, one can rewrite the
first term in r.h.s. of Eq.~(\ref{BG1}) as
\BEQ
\label{BG2}
\int \d \vec{r}~\nu (r)~\d \vec{v}
\left [
f\ln \frac{f}{ f_0}+f_0\ln \frac{f_0}{ f}
\right ]\ge 0,
\EEQ
since 
\BEQ
\int \d \vec{v}~f\ln \frac{f}{ f_0}\ge 0
\EEQ
for any distributions $f$, $f_0$.
The second term term in r.h.s. of Eq.~(\ref{BG1}) can be presented as
an integral over the corresponding surface in the coordinate space
\BEQ
\label{BG3}
\oint \d \vec{S}~\int\d \vec{v}~\vec{v}f\ln f
\EEQ
If there are no global currents this integral is zero, and we obtain
the H-theorem:
\BEQ
\label{BG4}
\dot{S}\ge 0
\EEQ
Monotonically increasing at finite times $S$ will saturate in the 
infinite-time limit: $f=f_0$. Substituting this expression in l.h.s of
the kinetic equation we will get the Gibbs distribution 
\BEQ
\label{BG5}
T={\rm const.},\qquad 
\label{BG6}
n(\vec{r})=\frac{1}{Z}e^{-\beta \phi (\vec{r})}
\EEQ
If there are global currents in the system, then the
actual distribution is not known a priori, and the full kinetic equation
has to be solved.

\subsection{Kinetic equation in spherical coordinates}
Since we will consider non-equilibrium distribution functions, which are
in a sense close to be isotropic, it is reasonable to employ spherical 
coordinates:
\begin{eqnarray}
\label{tamash1}
&& x = r\sin \theta \cos \varphi , \qquad
 y = r\sin \theta \sin \varphi , \qquad
 z = r\cos \theta , 
\end{eqnarray}
In the local Cartesian base $\{ \vec{e}_r$, 
$\vec{e}_{\theta}$, $\vec{e}_{\varphi} \}$ (e.g., $\vec{e}_r$ is a unit
vector, which points out from $r$ to $r+\d r$) velocity will have 
components
\BEQ
\label{tameshit3}
\vec{v}=
v_r \vec{e}_r + v_{\theta} \vec{e}_{\theta} +v_{\varphi} \vec{e}_{\varphi},
\EEQ
where
\begin{eqnarray}
\label{tamash2}
&& v_x = v_r\sin \theta \cos \varphi 
        +v_{\theta}\cos \theta \cos \varphi - v_{\varphi}\sin \varphi, \\
&& v_y = v_r\sin \theta \sin \varphi 
        +v_{\theta}\cos \theta \sin \varphi + v_{\varphi}\cos \varphi, \\
&& v_z = v_r\cos \theta - v_{\theta}\sin \theta, 
\end{eqnarray}
In the spherical coordinates the kinetic equation will read
\begin{eqnarray}
\label{tamash5}
\frac{\partial f}{\partial t}&&+v_{r}\frac{\partial f}{\partial r }
+\frac{ v_{\theta }  }{ r } \frac{\partial f}{\partial \theta }
+\frac{ v_{\varphi }  }{ r\sin \theta }\frac{\partial f}{\partial \varphi }
+\left [ \frac{v_{\theta }^2 + v_{\varphi }^2 }{r}
-\frac{\partial \phi}{\partial r }
\right ]
\frac{\partial f}{\partial v_r }
\nonumber\\
&&+\frac{ {\rm cot}\theta   }{ r }\left [
 v^2_{\varphi }\frac{\partial f}{\partial v_{\theta} } 
-v_{\varphi }v_{\theta }\frac{\partial f}{\partial v_{\varphi} }
\right ]
-\frac{ v_r   }{ r }\left [
 v_{\theta }\frac{\partial f}{\partial v_{\theta} } 
+v_{\varphi }\frac{\partial f}{\partial v_{\varphi} }
\right ] 
= {\cal C}_{BGK}[f]
\end{eqnarray}
We will consider non-equilibrium states with radial symmetry.
We will assume that 
$\phi (r,\theta , \varphi ) = \phi (r) $, and
the distribution function $f$ is
non-equilibrium exclusively through its radial properties:
\BEQ
\label{tamash6}
f(r,\theta , \varphi, v_r,  v_{\theta }, v_{\varphi } )=
f(r,v_r)\frac{1}{2\pi T(r)}\exp \left [  
-\frac{v_{\theta }^2+ v_{\varphi }^2 }{2T(r)}\right ],
\EEQ
namely there is no dependence on the angular coordinates, and the
angular velocities have relaxed to equilibrium.
In fact, Eq.~(\ref{tamash6}) is the minimal Ansatz, which allows to
study non-equilibrium states within the simplest spherical geometry.

For the collision term one has

\begin{equation}
\label{a42}
{\cal C}_{BGK}[f]
=-\nu (r)
\left [ f(r,v_r)-f_0(r,v_r) \right ]\times
\frac{1}{2\pi T(r)}\exp \left [  
-\frac{v_{\theta }^2+ v_{\varphi }^2 }{2T(r)}\right ],
\end{equation}
\begin{equation}
\label{a43}
f_0(r,v)=\frac{n(r)}{[2\pi T(r) ]^{1/2}}
\exp \left ( -\frac{(v-u)^2}{2T(r)}    \right ),
\end{equation}
where the mean radial velocity $u$ describes an average motion
(current of particles) along the radial direction. 

Let us now substitute Eq.~(\ref{tamash6}) to Eq.~(\ref{tamash5}), and 
integrate by $v_{\theta }$, $v_{\varphi }$.
The result will read
\BEQ
\label{tamash8}
\frac{\partial f}{\partial t}+v_{r}f'- \phi '\frac{\partial f}{\partial v_r }
+\frac{ 2T(r)   }{ r }\left [ f(r)\frac{v_r}{ T(r) } 
+\frac{\partial f}{\partial v_r }
\right ]
= - \ub (r)\left [ f(r,v_r)-f_0(r,v_r) \right ]
\EEQ
where $f'=\partial f(r,v_r)/\partial r$, $\phi '=\d \phi /\d r$.

\subsection{Moments}
Eq.~(\ref{tamash8}) is still complicated integro-differential equation.
To subtract the relevant information from it
one considers the subtracted and non-subtracted  moments
\begin{equation}
\label{a44}
M_{l}(r)=\int \d v_r~ (v_r-u)^{l}f(r,v_r),
\end{equation}
\begin{equation}
\label{a44x}
{\cal M}_{l}(r)=\int \d v_r~ v_r^{l}f(r,v_r),
\end{equation}
These two types of moments correspond to different experimental situations, 
since $M_{l}(r)$ are measured in the comoving frame, whereas ${\cal M}_{l}$
at the rest. We will use them in parallel choosing the one or the another
as appears most convenient.

The corresponding equations read:
\begin{eqnarray}
\label{a46}\dot M_{l}&+&l\dot{u}M_{l-1}+
\frac{1}{r^2}[r^2(M_{l+1}+uM_l)]'+l\left (\phi ' -\frac{2T}{r} \right ) M_{l-1}
+lu'[M_{l} + uM_{l-1} ]\nn\\
&=&-\ub (r) [ M_{l}  - \omega (l)~ nT^{l/2}~(l-1)!! ],
\end{eqnarray}
where $\omega (l)$ is zero for $l$ odd, and equal to $1$ for $l$ even.

\begin{eqnarray}
\label{a46x}\dot {\cal M}_{l}&+&
\frac{1}{r^2}(r^2{\cal M}_{l+1})'+l\left (\phi ' -\frac{2T}{r} \right ) 
{\cal M}_{l-1}\nn\\
&=&-\ub (r) [ {\cal M}_{l}  - ~ n\sum _{p=0}^l
u^{l-p}T^{p/2}~\omega (p)(p-1)!! ],
\end{eqnarray}

Recall that, by construction, the low moments 
\begin{eqnarray}
\label{a47a} &M_{0}=n(r), 
\qquad M_{1}=0, 
\qquad & M_{2}=n(r)T(r), \\
\label{a47d} &{\cal M}_{0}=n(r), 
\qquad {\cal M}_{1}=n(r)u(r), 
\qquad &{\cal M}_{2}=n(r)[T+u^2], 
\end{eqnarray}
are the same for $f_0(r,v_r)$ and $f(r,v_r)$. 
Since this fact expresses conservation
of probability, momentum and energy, the equations
(\ref{a46}, \ref{a46x}) with $l=0,1,2$ will be universal,
namely they do not depend on the concrete form of the
collision integral.
Notice also that the third moments are connected as:
\BEQ
\label{kozak}
{\cal M}_3=M_3+nu^3+3unT.
\EEQ
Here ${\cal M}_3$ is energy current, which consists of heat current
$M_3$, transfer of kinetic energy $nuu^2$, and work done by pressure
$3unT=3uP$.

Let us write down the first members of Eq.~(\ref{a46})
\begin{eqnarray}
\label{a48a} l=0:\, && r^2\dot n(r)+(r^2n\,u)'=0, \\
\label{a48b} l=1:\, && n\dot{u}+(nT)'+n\phi '+nuu' =0, \\ 
\label{a48c} l=2:\, 
&&n\dot T+\frac{1}{r^2}{(r^2M_{3})'}+2u'nT+unT'=0, 
\\
\label{a55a}
l=3:\, && \dot M_3+3nT\dot{u}+\frac{1}{r^2}[r^2 (M_{4}+uM_3)]'+
3u'[M_3+nuT]\nn \\ &&+3nT(\phi '-\frac{2T(r)}{r})=-\ub (r)M_{3},
\\
\label{a55b} 
l=4:\, &&\dot M_4+4M_3\dot{u}+\frac{1}{r^2}[r^2 (M_{5}+uM_4)]'+
4u'[M_4+uM_3]\nn \\ &&+4M_3(\phi '-\frac{2T(r)}{r})
=-\ub (r)(M_{4} - 3nT^2)
\end{eqnarray}
Eq. ~(\ref{a48a}) expresses conservation of the mass inside 
the sphere $r$. When $u=0$ Eq.~(\ref{a48b}) expresses 
hydrostatic equilibrium; when $u\neq0$ it contains flow terms, that
should be easy to understand.
The equation for $l=2$ expresses the conservation of energy; for
obtaining it we have also inserted Eq. (\ref{a48a}).

Somewhat different, but of course, totally equivalent expressions
will be obtained for the ${\cal M}$-moments.

Eqs.~(\ref{a48a}-\ref{a55b}) are inherently attached to the Poisson
equation:
\BEQ
\label{poisson}
\frac{1}{r^2}(r^2\phi ')'=4\pi Gn(r).
\EEQ
The density $n(r)$ should satisfy to the condition of integrability,
\BEA
4\pi\int_0^{\infty}\d r\, n(r)r^2=N.
\EEA

Notice that equations with $l\ge 2$ can be written in a form, which does
not contain $\phi '$ explicitly. In that respect their combination
can be viewed as equations of state. For example, using Eq.~(\ref{a48b}),
the for result for $l=3$ reads 
\BEQ
\label{khazar}
\dot M_3+\frac{1}{r^2}[r^2 (M_{4}+uM_3-3nT^2)]'+
3u'M_3+3nTT' =-\ub (r)M_{3}.
\EEQ

\section{Steady state.}
As was already indicated in the introduction, we will consider the
stationary (steady) situation.
This case, where all quantities are time-independent, is worth 
to study for its own sake, since there are clear observational
evidences that typical globular clusters do have such a phase in their
evolution \cite{shapiro,MeylanHeggie}. On the other hand, it is a necessary
step towards understanding of more general, time-dependent situation.

In the situation, where the known conditions for equilibrium are satisfied
(e.g., absence of energy and/or probability currents), the
self-gravitating system under study has the Gibbs distribution as its
only steady state. This point was already illustrated by us when
considering the H-theorem in section \ref{H}.
In certain range of temperatures this distribution is physical (e.g.,
it is at least metastable), but it loses its stability for
temperatures lower than a certain critical temperature  \cite{paddy}.
This is the notorious phenomenon of gravo-thermal instability
(collapse) \cite{spitzer,saslaw,paddy}. However, there are situations,
where one expects that the very reasons for the existence of the
equilibrium can be invalid. The most typical situation of that kind
appears with binary star formation and tightening processes in the central
region of a star cluster \cite{spitzer,saslaw,shapiro}. These processes are
accompanied with a release of energy, so that the rest of the system
(namely its outer with respect to a relatively small part, where the
binary-formation process occurs) can be considered as being subjected
to sources of energy put in the central part. The existence of energy 
currents from the central part leads to a non-equilibrium steady
state.
Additionally, they can stabilize the behavior of the self-gravitating system
preventing its collapse, since the mechanism of the gravothermal
instability is connected with spontaneous energy currents towards the
center \cite{spitzer,saslaw,shapiro}. 

A somewhat more exotic example of a non-equilibrium steady state can be
provided by the existence of a black hole in the central part of a
star cluster \cite{spitzer,saslaw,shapiro}. 
Then one has a stationary current of consumed stars, which is directed
towards the center, as well as a related current of energy.

Our setup with the moments equations is especially suitable for
describing the above non-equilibrium states, since there are explicit
expressions for the energy current, ${\cal M}_3$ and the current of
particles $u$.

\subsection{Ideal hydrodynamics}
This scheme is determined by a condition $M_3=0$ (no heat current).
The nonequilibrium character is displayed only through the
stationary current of particles $u$.

Eqs.~(\ref{a48a}-\ref{a48c}) read
\begin{eqnarray}
\label{aa48a}  && (r^2n\,u)'=0,  \qquad
\label{aa48b}  (nT)'+n\phi '+nuu' =0, \qquad 
\label{aa48c}  2u'nT+unT'=0.
\end{eqnarray}
Recall that
the usual gibbsian equilibrium is a particular case of this
situation, which is realized for $u=0$, and an additional assumption
that $T={\rm const}$. In this case
Eqs.~(\ref{aa48a}, \ref{aa48c}) are trivial, and Eq.~(\ref{aa48b}) leads
to the Gibbs distribution. 

Let us first assume that $u$ does not depend of $r$. Then
Eq.~(\ref{aa48c}) predicts that $T={\rm const}$. With 
Eq.~(\ref{aa48a}) one gets
\BEQ
\label{ura}
n=\frac{c}{u}\,\frac{1}{r^2},\qquad c={\rm const},
\EEQ
Notice that similar to the isothermal case
this distribution is non-integrable at infinity, and therefore
predicts infinite mass.
Being combined with the Poisson equation (\ref{poisson}),
Eq.~(\ref{aa48a}) offers an exact solution:
\BEQ
\label{ura1}
\phi =2T\ln r +\phi _0.
\EEQ

To study the case $u\not = {\rm const}$ one 
first notice that Eq.~(\ref{aa48c}) can be exactly integrated
\BEQ
\label{ura2}
T=\frac{c_1}{u^2},\qquad c_1={\rm const}
\EEQ
Eqs.~(\ref{ura}, \ref{ura2}) combined with (\ref{aa48a}-\ref{aa48c})
lead to
\begin{eqnarray}
\label{lo1}  && c_1u\,r^2(u^{-3}r^{-2})'\phi '+u'u=0 ,  \qquad
\label{lo2}  u(r^2\phi ')' =4\pi Gc.
\end{eqnarray}
To study self-similar solutions, one 
makes an Ansatz $u= u_0\,r^{\alpha}$, and gets from Eq.~(\ref{lo1})
\BEQ
\label{ura3}
-c_1u_0^{-2}(3\alpha +2)r^{-2\alpha -1}+\phi '+\alpha\, u_0^2 \,r^{2\alpha -1}=0
\EEQ
Consulting with the Poisson equation one observes that for small distances
the last term in Eq.~(\ref{ura3}) can be neglected, and as the result
one has the following asymptotic solution at small distances $\alpha=1/2$ 
and 
\BEA
\label{ura4}
\phi '\simeq\frac{7c_1}{2u_0^2}\, r^{-2}+\frac{8\pi Gc}{u_0}\,r^{-3/2},
\qquad 
\label{ura6} \qquad 
T\simeq \frac{c_1}{u_0^2}\, r^{-1}, \qquad 
\label{ura7}
\qquad n\simeq  \frac{c}{u_0}\, r^{-5/2}.
\EEA
It is seen that the density is singular at the origin, but still 
integrable. The first term in $\phi'$ indicates
the presence of the central mass $7c_1/(2\,u_0^2\,G)$. This implies
the following choice: $u_0<0$, $c<0$, and then the obtained solution
describes a stationary consumption regime of stars by the central
mass. The energy current ${\cal M}_3=nu^3+3nuT$ is also directed
towards the center.

\subsection{Overdamped situation}
Starting since this moment, we will into the consideration an energy
current, described by $M_3$. We make a self-consistent assumption that
there is a range of distances close to the inner part (i.e., the
core), where collisions dominate. In this overdamped regime
$\eta =({\rm dynamic\, time})/({\rm kinetic \, time}) $ 
is large, and the large $\ub (r)$ in the r.h.s. of 
Eqs.~(\ref{a55a}, \ref{a55b}) imposes the Gaussian 
behavior for the corresponding moments:
\begin{eqnarray}
\label{a49}
&& M_{3}={\cal O}(1/\eta ),\qquad
 M_4-3nT^2={\cal O}(1/\eta )
\label{a499}
\end{eqnarray}
Using the last result as $ M_4\simeq 3nT^2$ one gets from Eq.~(\ref{a55a}) 
\BEA 
3n(r)T(r)T'(r) =-\ub (r) M_3(r) 
\label{M3}
\EEA
as a constitutive equation for the overdamped case. Let us consider separately 
cases with $u=0$, and $u\not =0$.

\centerline{\it The case: $u=0$}

Then Eq.~(\ref{a48c}) reduces to 
\BEQ
\label{bora1}
M_3r^2=\nu _3={\rm const}
\EEQ
If $\nu_3>0$ there is a solution
\BEQ
T=\left(\frac{2\gamma\nu_3}{21}(\frac{1}{r}+\frac{1}{r_0})\right)^{2/7}
\EEQ
with some integration constant $r_0$.
The condition $\nu_3>0$ means a positive outward flow of energy.
One thus gets for small $r$
\BEQ 
T\simeq T_0r^{-2/7}, \qquad T_0=\left(\frac{2\gamma\nu_3}{21}\right)^{2/7}.
\EEQ
Taking this into account one easily gets from the Poisson equation and 
hydrostatic equilibrium
\BEQ n(r)\simeq n_0 r^{-16/7}, \qquad 
n_0=\frac{1}{4\pi G}\frac{90}{49}\, T_0, \qquad
\frac{16}{7}=2.2857143, \EEQ
\BEQ \phi '(r)\simeq \phi _0 r^{-9/7}, \qquad 
\phi _0=\frac{18}{7}T_0.
\label{kuvald}
\EEQ
Notice that density is singular at the origin, but integrable.
Thus, this solution can be intended to describe a stationary regime
without a central mass (as follows from Eq.~(\ref{kuvald})), but with
a stationary tightening of binaries at the very central region. In our
setup this last process is reflected through a constant outward heat
current. 

The mass insider a sphere of radius $r$ scales as
$M(r)\sim r^{5/7}$. The inverse is $r\sim M^{7/5}$.
These are so-called Lagrange radii. If one chooses a set of
equidistant $M$ values, e.g. $M=i/20$, $i=1,\cdots,20$, then
the radii will be closer to each other near the center than
in the outer parts of the cluster. This happen one since $16/7>2$.
If, on the other hand, the density would be finite at the center,
then one would have $r\sim M^{1/3}$, implying large separations
between Lagrange radii near $M=0$.

\centerline{\it The case: $u\not = 0$}
\label{evapo}
Here one can still use Eq.~(\ref{M3}) for $M_3$, which now does not reduce
to ${\rm const}/r^{2}$. Having substituted this equation to Eq.~(\ref{a48c}),
one obtains
\begin{eqnarray}
\label{napal1} l=0:\, && (r^2un) '=0, \qquad
\label{napal2} l=1:\,  (nT)'+n\phi '+u'un=0, \\ 
\label{napal3} l=2:\, 
&& -\frac{3}{\gamma}\frac{1}{r^2}[r^2T^{5/2}T']'+unT'+2u'nT=0 , 
\end{eqnarray}
For small distances the self-consistent solution of these equations
reads
\BEA
&&n(r)\simeq n_0r^{-12/5},\quad
T(r)\simeq T_0r^{-2/5},\quad u(r)\simeq u_0r^{2/5},
\quad \phi '(r)\simeq \frac{14}{5}T_0r^{-7/5},
\label{kuvald1}
\EEA
where 
${6}T_0^{5/2}=5\gamma u_0n_0$.
Notice that there is no central mass (as follows from
Eq.~(\ref{kuvald1})). Moreover, $u_0$ has to be positive and 
$u(0)=0$. Therefore, 
the obtained solution can be suitable to describe evaporation phenomena.

\subsection{Underdamped situation}

There is yet another extremal situation realized at sufficiently large
distances from the center, where the behavior of the system is 
almost collisionless (underdamping).

\centerline{\it The case with $u=0$}

In the underdamping regime 
one expects that all moments with $k\geq 3$ will be equally 
important, therefore they have to behave in the nearly similar way.
Looking for a self-consistent, long-distance solution of the hierarchy
(\ref{a48a}-\ref{a55b}) one gets:
\BEA
\label{lolo1}
&&n(r)\simeq\frac{n_0}{r^{7/2}},\qquad
\label{lolo2}
T(r)\simeq\frac{T_0}{r},\qquad
\label{lolo3}
\ub (r)\simeq \gamma \frac{n_0T^{-3/2}_0}{r^2},\\
&&M_3=\frac{\nu _3}{r^2},\qquad
\label{lolo4}
M _4\simeq\frac{\nu _4}{r^2}+ \gamma\frac{\nu _3n_0T^{-3/2}_0}{r^3},
\EEA
where $\nu _4$, $n_0$, $T_0$ are constant. At this stage of our asymptotic 
analysis they will remain unfixed. 
For $\phi '$ one has from the Poisson equation:
\BEQ
\label{ole1}
\phi '(r)=\frac{GM}{r^2}-\frac{8\pi G n_0}{r^{5/2}},
\EEQ
where the first term is the standard asymptotic limit for a finite-mass 
cluster with $M$ as the total mass. The second term follows from 
Eq.~(\ref{lolo1}).
The behavior of $M _5$ is qualitatively 
the same as of $M _4$ as seen from Eq.~(\ref{a55b}).
Our expression for the density given by Eq.~(\ref{lolo1}) coincides
with that obtained in \cite{shapiro} for the halo of a globular
cluster by means of qualitative physical considerations.

Having substituted Eqs.~(\ref{lolo1}, \ref{lolo2}) to hydrostatic
equilibrium equation (\ref{a48b}), one 
gets the correction terms to the density and temperature
\BEQ
\label{lolo40}
n(r)=\frac{n_0}{r^{7/2}}\left [1 +\frac{n_1}{r^{1/2}}+    ...\right ],
\qquad 
\label{lolo5}
T(r)=\frac{T_0}{r}\left [1 +\frac{T_1}{r^{1/2}}+    ...\right ],
\EEQ\BEQ
\label{lolo6}
5T_0(n_1+T_1)=GM (n_1+8\pi G n_0)
\EEQ
with constant $n_1$, $T_1$.

\centerline{\it The case with $u\not =0$}

This case is capable to describe the evaporation     
phenomenon \cite{spitzer,saslaw} as we indicated 
already in section \ref{evapo}. Having this in mind, 
we will look for a solution, which has the Maxwellian 
structure for very long distances, since there it
is supposed to describe free escaped stars.
In other words, for those scales only first two moments 
$u=u_0>0$ and $T=T_0>0$ are nonzero, and should be viewed as boundary 
conditions. 
Possible solutions of this kind should be then matched with the 
overdamped solutions having finite $u$. Notice that the equation
(\ref{a48c}) for energy can be written in a simplified form
\begin{eqnarray}\label{khurshud}
\nu _3(r)'+\frac{J(r)}{u^2(r)}\,[u^2(r)T(r)]'=0 ,
\end{eqnarray}
where $M_3=\ub _3(r)\, r^2$, and where $J=r^2nu$ is the constant 
current of particles.

Searching again for a self-consistent solution, one gets
\BEA
\label{lolo7}
&&r^2n(r)\simeq n_0+\frac{n_{10}+n_{11}\ln r}{r}
+\frac{n_{20}+n_{21}\ln r+n_{22}(\ln r)^2}{r^2},\\
\label{lolo8}
&&T(r)\simeq T_0+\frac{T_{10}+T_{11}\ln r}{r}
+\frac{T_{20}+T_{21}\ln r+T_{22}(\ln r)^2}{r^2},\\
\label{lolo9}
&&r\phi '\simeq \phi _0+\frac{\phi _{10}+\phi _{11}\ln r}{r}
+\frac{\phi _{20}+\phi _{21}\ln r +\phi _{22}(\ln r)^2}{r^2},\\
\label{lolo10}
&&u(r)\simeq u_0+\frac{u_{10}+u_{11}\ln r}{r}
+\frac{u_{20}+u_{21}\ln r+u_{22}(\ln r)^2}{r^2},\\
\label{lolo11}
&&\nu _3(r)\simeq \nu _{30}+\frac{\nu _{310}+\nu _{311}\ln r}{r}
+\frac{\nu _{320}+\nu _{321}\ln \nu _{322}(\ln r)^2}{r^2},\\
\label{lolo12}
&&\ub (r)\simeq \gamma \frac{n_0T^{-3/2}_0}{r^2}.
\EEA
Higher-order terms can be presented analogously.
Notice that density does not have to be integrable at infinity
when considering stationary regimes of evaporation. Indeed, the only
 way to have a truly stationary evaporation regime is to
have an infinite amount of stars, the most of them being located at infinity.
The appearance of logarithms in the asymptotic expansions is mathematically
connected with non-integrable character of the density.

It is seen as well that the leading-order behavior of $\ub (r)$
is the same for both underdamped regimes (with and without evaporation).

At the present stage of our asymptotic analysis we able to obtain only
certain relations between the coefficients of (\ref{lolo7}-\ref{lolo12}).
\begin{eqnarray}
\label{kurban-2}
&&(r^2\phi ')'=4\pi Gr^2n(r)\,\to 
 \phi _0=4\pi Gn_0, \quad n_{11}= u_{11}=0, \quad
\phi _{11}=4\pi G n_{10}; \\
\label{kurban1}
&&unr^2=J,\,\to 
J=u_0n_0, \qquad n_0u_{10} + n_{10}u_{0}=0, \qquad
n_0u_{11} + n_{11}u_{0}=0;\\
\label{kurban5}
&&(nT)'+\phi '+\frac{n}{2}(u^2)'=0,\,\to 
 \phi _0=2T_0, \\
&& n_0\phi _{10}-2n_0u_0u_{10} = 3T_{10} n_{0}-3T_{11} n_{0} +2n_{10}T_0 , \qquad
3T_{11} =\phi_{11};\\
\label{kurban7}
&&\nu _3'+\frac{J}{u^2}[u^2T]'=0 ,\,  \Longrightarrow \, \nonumber\\
&& J(T_{11}-T_{10})-2T_0u_{10} +\nu _{311}+\nu _{310} = 0 , \qquad
-JT_{11} = \nu _{311}.
\end{eqnarray}
The equations support an additional simplification
$T_{10}=\phi _{10}=0$,
and consequently
$4\pi G n_0 = T_0 - 2u_0^2$. 
This equation connects the energetic characteristics of the considered 
solution.

\section{Conclusion}

This contribution was devoted to
the kinetic theory of many-body self-gravitating systems
in the framework of the model kinetic equation developed by
Bhatnager, Gross and Krook \cite{bgk}. The main advantage of using this
kinetic model (as well as other, more refined kinetic models) is
that it produces rather tractable schemes in contrast to the full
kinetic equations, which are typically quite difficult to
investigate. On the other hand, model kinetic equations offer rather
sensible results, which are usually at least in a qualitative
agreement with those obtained from the full Boltzmann
kinetic equation. The physical reason of this success is that a good
amount of statistical phenomena are quite insensitive to the details
of the collision integral. In order  to explain them it is 
enough to use the simplest relaxation mechanism which only takes into 
account necessary conservation laws of energy, momentum and
probability (number of particles) and the correct characteristic 
relaxation time.

The assumptions on the spherically symmetric character of the
considered system and the assumed radial character of non-equilibrium 
led us
to the kinetic equation (1), which was then studied by the method
of moments. The straightforward analytic investigation allowed us to
identify several self-similar regimes of a steady but non-equilibrium
behavior. These solutions corresponds to the presence of the central
black hole in the system (section 3.1), the influence of binary-formation
and tightening processes to the system (section 3.2) and evaporation
phenomena (section 3.3). In this way we reproduced certain known results
\cite{shapiro} and also obtained a number of new regimes of behavior. 

We believe that the model kinetic approach will be potentially quite
powerful in application to many-body, self-gravitating systems. We plan
in future to turn to the dynamical aspects of this approach, as well as
construct more realistic (from the viewpoint of applications in
astrophysics) kinetic models, that e.g account for the strong 
anisotropy of the velocity distribution in the outer part of the
cluster.

\end{document}